\documentclass[twocolumn,twoside,letterpaper,11pt]{article}
\usepackage{graphicx,xrrc,natbib}
\usepackage{times}

\def\spose#1{\hbox to 0pt{#1\hss}}
\def\ltsimm{\mathrel{\spose{\lower 3pt\hbox{$\sim$}}
        \raise 2.0pt\hbox{$<$}}}
\def\gtsimm{\mathrel{\spose{\lower 3pt\hbox{$\sim$}}
        \raise 2.0pt\hbox{$>$}}}
\def\Mdot{\hbox{${\dot M}$}}

\def\km{{\rm\thinspace km}}
\def\cm{{\rm\thinspace cm}}
\def\s{{\rm\thinspace s}}
\def\yr{{\rm\thinspace yr}}
\def\g{{\rm\thinspace g}}
\def\K{{\rm\thinspace K}}
\def\mG{{\rm\thinspace mG}}
\def\kmps{\hbox{${\rm\km\s^{-1}\,}$}}

\def\erg{{\rm\thinspace erg}}
\def\Hz{{\rm\thinspace Hz}}

\def\keV{{\rm\thinspace keV}}
\def\ster{{\rm\thinspace ster}}
\def\ergps{\hbox{${\rm\erg\s^{-1}\,}$}}
\def\Msol{\hbox{${\rm\thinspace M_{\odot}}$}}

\def\Rsol{\hbox{${\rm\thinspace R_{\odot}}$}}
\def\Msolpyr{\hbox{${\rm\Msol\yr^{-1}\,}$}}
\def\pcm{\hbox{${\rm\cm^{-1}\,}$}}
\def\pcm3{\hbox{${\rm\cm^{-3}\,}$}}
\def\ergpscm3Hz{\hbox{${\rm\ergps\cm^{-3}\Hz^{-1}\,}$}}
\def\ergpscm3Hzster{\hbox{${\rm\ergps\cm^{-3}\Hz^{-1}\ster^{-1}\,}$}}
\def\gpcm3{\hbox{${\rm\g\cm^{-3}\,}$}}
\def\ergpcm2{\hbox{${\rm\erg\cm^{-2}\,}$}}
\def\ergpcm3{\hbox{${\rm\erg\cm^{-3}\,}$}}
\def\wr{{\rm WR\thinspace}}

\def\fdg{\hbox{$.\!\!^\circ$}}

\def\farcs{\hbox{$.\!\!^{\prime\prime}$}}

\begin{document}


\title{X-RAY AND RADIO EMISSION FROM \\
COLLIDING STELLAR WINDS}


%
%
%
%


\author{    J. M. Pittard                         } 
\institute{ School of Physics and Astronomy, The University of Leeds         } 
\address{   Woodhouse Lane, Leeds, LS2 9JT, U.K.        } 
\email{     jmp@ast.leeds.ac.uk                         } 

\author{    S. M. Dougherty, R. F. Coker, M. F. Corcoran}
\email{     Sean.Dougherty@hia.nrc.ca, robc@lanl.gov, corcoran@barnegat.gsfc.nasa.gov}


\maketitle

\abstract{The collision of the hypersonic winds in early-type binaries
produces shock heated gas, which radiates thermal X-ray emission, and
relativistic electrons, which emit nonthermal radio emission. We
review our current understanding of the emission in these spectral
regions and discuss models which have been developed for the
interpretation of this emission. Physical processes which affect
the resulting emission are reviewed 
and ideas for the future are noted.}

\section{Introduction}
It is well established that massive O-type and
Wolf-Rayet (WR) stars have powerful, highly supersonic winds with
high mass-loss rates and terminal speeds. These stars are all
hot and highly luminous for their mass and their winds are driven by
momentum transferred from the radiation field by line scattering.
Representative values for some of the defining parameters of massive
stars are noted in Table~\ref{tab:massive_properties}.

In addition, a high proportion of massive stars are in binary and
multiple systems; a volume-limited sample found more than $40$\% of WR
stars in multiples \citep[][]{vdH:2001}. A variety of phenomena related
to the collision of the two winds can be observed. For instance,
infrared (IR), optical and ultraviolet spectra of many WR binaries show
phase-locked variability in P-Cygni and flat-topped emission lines
\citep[e.g.,][] {Wiggs:1993,Rauw:1999,Stevens:1999,Hill:2000}; these
are signatures of wind-wind collisions. Dust formation may also occur
in the wind collision region (WCR) and is observed as an IR
excess. Binary systems which exhibit dust formation are categorized
either into persistent or episodic dust makers \citep[see][for a
review]{Williams:2002}.

Excess X-ray emission from shock-heated plasma produced in the
WCR is another wind signature -- in some systems the X-ray
luminosity is two orders of magnitude higher than suggested by the
canonical relationship, $L_{\rm x}/L_{\rm bol} \sim 10^{-7}$, from
early-type stars \citep{Chlebowski:1991,Moffat:2002}. Early-type
binaries are also found to be stronger X-ray sources than their single
counterparts in a statistical sense
\citep{Pollock:1987,Chlebowski:1991}. Massive binaries often display
phase-locked orbital X-ray variability resulting from the changing
line of sight into the system
\citep{Willis:1995,Corcoran:1996}. Recent {\it Chandra} observations
indicate that the X-ray emission is nonisothermal, that the lines are
in general narrow, and that there is strong emission from forbidden
lines \citep{Corcoran:2001a, Pollock:2004}.

Finally, nonthermal radio emission from WR stars is also a good
indicator of wind-wind interaction \citep[e.g.,][]{Dougherty:2000b}.
In binary systems the strong shocks formed by the collision of the two
stellar winds are natural sites for particle acceleration.  Spatially
resolved emission from the WCR in several nearby WR+OB binaries,
including \wr140 (Tony Beasley, private comm.),
\wr146 \citep{Dougherty:1996,Dougherty:2000a}, and \wr147
\citep{Moran:1989, Churchwell:1992, Williams:1997, Niemela:1998}, has
confirmed this picture.

\begin{table*}[ht]
  \begin{center}
  \caption{\small Parameters of different evolutionary stages during the life of
massive stars. Very massive stars ($M \gtsimm 60 \Msol$) are
thought to evolve from an O-type main sequence star, through a
luminous blue variable (LBV) phase to a Wolf-Rayet (WR) star, before
ending their lives in a supernova (or hypernova) explosion. These
stages represent the transition from core-H burning to shell-H burning
to core-He burning.}\medskip
  \label{tab:massive_properties}
  \begin{tabular}{llll}
    Phase & Lifetime (yrs) & $\Mdot (\Msolpyr)$ & $v_{\infty} (\kmps)$\\
    \hline\hline
    O & $\sim$few $\times 10^{6}$ & $\sim 10^{-7} - 10^{-5}$ & $\sim 1000-3000$ \\
    LBV & $\sim 10^{4}$ & $\sim 10^{-3}$ (+outbursts) & $\sim 300-500$ \\
    WR & $\sim$few $\times 10^{5}$ & $\sim 10^{-5} - 10^{-4}$ & $\sim 1000-4000$ \\
    \hline
  \end{tabular}
  \end{center}
\end{table*}

In this paper we focus on the theoretical modeling of
X-ray and radio emission from colliding wind binaries (CWB).
X-ray observations provide a direct probe of the conditions within the
wind-wind collision zone and of the unshocked attenuating
material in the system \citep[e.g.,][]{Stevens:1992,Pittard:1997}.
Analysis of the X-ray emission can determine the mass-loss
rates of the stars and the speeds of the stellar winds. It is also possible
to quantify various processes which occur in
massive stellar binaries and high Mach number shocks.
In contrast, the analysis of nonthermal radio emission from CWB systems 
can yield estimates of the efficiency of particle acceleration and
the equipartition magnetic fields at the surface of the stars, and 
the mass-loss rates and clumping fractions of the stellar winds. 
CWB systems are also excellent objects for investigating the
physics of diffusive shock acceleration as their
geometry is in many cases very simple, and because they provide access
to a higher density regime than is possible through studies of
supernova remnants. Systems with highly eccentric orbits which show 
large modulation of the nonthermal emission, such as \wr140 
\citep{Williams:1990, White:1995}, should be
especially useful in this regard.

\begin{figure}[h]
  \begin{center}
    \includegraphics[width=\columnwidth]{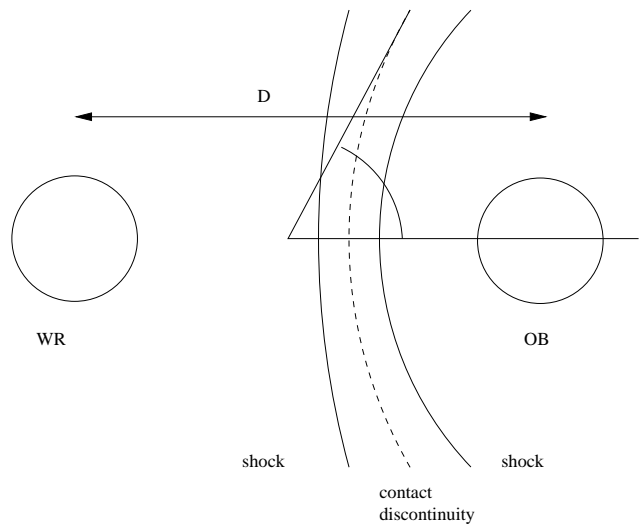}
    \caption{\small Schematic figure of the geometry of the wind-wind 
collision in a wide binary. For the purposes of this work we assume that
the binary consists of a WR and OB-star, though other combinations (such as
O+O-star) also occur. Since the binary is assumed to be wide the winds
collide after they have reached their terminal speeds, and the distortion
of the collision region due to the orbital motion of the stars is not
visible until far downstream. The half-opening angle is indicated. Note that the stars are enlarged for clarity.}
    \label{fig:cw_schematic}
  \end{center}
\end{figure}

This paper is organized as follows.
In Section~\ref{sec:geom} we briefly review the basic geometry of the
wind-wind interaction. In Section~\ref{sec:xray} we review the current state of
the theoretical models used to analyze the X-ray emission from
colliding wind systems. Systems where the cooling lengths may be resolved are
amenable to study with hydrodynamical codes, and have traditionally been
the focus of work in this area. However, a recent development in the 
modeling of highly radiative systems is reported.
Models which focus on the radio emission are reviewed
in Section~\ref{sec:radio}, with particular attention paid to the first
hydrodynamical-based investigation which includes thermal and nonthermal
processes. Section~\ref{sec:other} contains a brief review of other
mechanisms related to CWBs, and in
Section~\ref{sec:summary} we summarize and conclude.

\section{Geometry of the wind-wind collision}
\label{sec:geom}
Consider the wind-wind collision in a wide binary as shown in the schematic in
Fig.~\ref{fig:cw_schematic}. The winds collide at high speeds ($v \sim
1000-3000 \kmps$) creating a region of high temperature shocked gas
between them ($T \sim 10^{7}-10^{8} \K$). The position of the
WCR is determined by ram-pressure balance requirements,
and the shocked winds are separated by a contact discontinuity. If the winds
are of equal strength the contact discontinuity will be a plane midway
between the stars, but in the more general case the
WCR wraps around the star with the weaker wind. 
Maximum values for the density, pressure,
and temperature occur at the stagnation point, a mathematical singularity
which occurs where the contact discontinuity intercepts the line
between the centers of the stars. If the
winds are spherical the distances $r_{\rm WR}$ and $r_{\rm OB}$ from
the WR and OB-stars, respectively, to the stagnation point are
\begin{equation}
r_{\rm WR} = \frac{1}{1 + \eta^{1/2}} D, \hspace{10mm}
r_{\rm OB} = \frac{\eta^{1/2}}{1 + \eta^{1/2}}D,
\label{eq:rob}
\end{equation}
where the wind momentum ratio, 
\begin{equation}
\eta = \frac{\Mdot_{\rm OB} v_{\rm OB}}{\Mdot_{\rm WR} v_{\rm WR}},
\label{eq:eta}
\end{equation}
$D$ is the stellar separation, $\Mdot_{\rm WR}$ and $\Mdot_{\rm OB}$
are the mass-loss rates of the WR and OB stars, and $v_{\rm WR}$ and
$v_{\rm OB}$ are their wind speeds. From the values in
Table~\ref{tab:params} we see that the dimensionless parameter $\eta$
is usually small i.e. the WCR is nearer to the OB star than to the WR
star.

This simple picture is a good approximation in wide binaries, where
the winds are accelerated to their terminal speeds before they
collide. The half opening angle is the angle between the
line-of-centers through the stars and the asymptotic direction of the
contact discontinuity (in the absence of orbital motion). In binaries
where the winds achieve terminal speeds the half-opening angle of the WCR
(specifically the contact discontinuity) approaches an asymptotic
value which is approximated by
\begin{equation}
\label{eq:halfopen}
\theta \approx 2.1 \left(1 - \frac{\eta^{2/5}}{4}\right)\eta^{1/3},
\end{equation} 
for $10^{-4} \leq \eta \leq 1$ \citep[see][]{Eichler:1993}.  The angle
between the contact discontinuity and each shock is largest when the
hot gas cools primarily through adiabatic expansion as it flows out of
the system (hereafter ``adiabatic systems''), and approaches zero if
the WCR is able to cool efficiently (hereafter ``radiative systems'').

Whether the system is radiative or adiabatic can be determined by the
characteristic cooling parameter \citep[see][]{Stevens:1992}
\begin{equation}
\label{eq:chi}
\chi = \frac{t_{\rm cool}}{t_{\rm dyn}} \approx 
\frac{v_{8}^{4} D_{12}}{\Mdot_{-7}},
\end{equation}
where $t_{\rm cool}$ and $t_{\rm dyn}$ are characteristic timescales
for cooling and for the flow dynamics, $v_{8}$ is the wind speed in
units of $1000 \kmps$, $D_{12}$ is the stellar separation in units of
$10^{12} \cm$, and $\Mdot_{-7}$ is the mass-loss rate of the star in
units of $10^{-7} \Msolpyr$. Here we have assumed that the temperature
of the shocked gas correspond roughly to the minimum in the cooling
function at $T \sim 10^{7}\K$ (see Fig.~\ref{fig:xray_ingredients} for
sample cooling functions).

Values of $\chi$ and other interesting parameters for some of the most
well-studied systems are listed in Table~\ref{tab:params}.  It is
clear from Table~\ref{tab:params} that CWBs are a
very diverse class of objects, and this is reflected in the wide range
of phenomena that occur in them.

\begin{table*}[ht]
  \begin{center}
  \caption{Parameters of interest for some of the most important
colliding wind systems. The systems are ordered according to their
orbital period. Where the orbit is eccentric a range in the possible
values is noted. The WCR may crash onto the O-star photosphere in V444~Cyg,
which makes an estimate for $\chi_{\rm OB}$ somewhat difficult in this 
system.}\medskip
  \label{tab:params}
  \begin{tabular}{llllll}
    System & Orbital period (d) & Separation (AU) & Density ($\pcm3$) & $\chi_{\rm WR}$ & $\chi_{\rm OB}$\\
    \hline\hline
    WR~139 (V444~Cyg) & 4.2 & 0.2 & $\sim 10^{10}$ & $\ll 1$ & ? \\
    WR~11 ($\gamma^{2}$~Vel) & 78.5 & 0.8-1.6 & $\sim 10^{9}$ & $\sim 0.5-1$ &
       $250-500$ \\
    WR~140 & 2899 & $\sim 1.9-30.7$ & $\sim 10^{9} - 10^{7}$ & $\sim 2-50$ &
       $100-1000$ \\
    WR~147 & $> 10^{5}$ & $>410$ & $\leq 10^{4}$ & $> 30$ & $>1000$ \\
    \hline
  \end{tabular}
  \end{center}
\end{table*}

\section{X-ray emission from CWBs}
\label{sec:xray}
\citet{Prilutskii:1976} and \citet{Cherepashchuk:1976} proposed that
binary star systems, in which one or both components are massive OB or
WR stars with strong stellar winds, should emit a detectable flux of
X-rays as a result of the shocked gas formed where the winds collide.

We see from Eq.~\ref{eq:chi} that since $\chi \propto D$, short-period
systems tend to be radiative, while wide systems tend to be adiabatic.
In highly radiative systems ($\chi \ll 1$), the X-ray luminosity $L_{x}
\propto \Mdot v^{2}$. If both of the shocked winds satisfy $\chi \ll 1$
then the faster wind will dominate the emission \citep{Pittard:2002b}
-- otherwise, the wind with the lower value of $\chi$ will in general
dominate $L_{x}$. Alternatively, if both of the shocked winds are largely
adiabatic, then $L_{x} \propto \Mdot^{2}/v^{3.2}D$
\citep{Stevens:1992}, and the stronger wind dominates the emission.
In this case, $\geq 90\%$ of the X-ray luminosity is emitted within an
off-axis distance equal to $D$ \citep[see][and references
therein]{Pittard:2002b}.

The particular method used to construct a theoretical model of the
X-ray emission from a CWB is to a large extent determined by the
nature of the WCR. Most work to date has focused on systems where the
stellar separation is fairly large and where the WCR
is largely adiabatic. In such systems it is normal to use a
hydrodynamical model to obtain a description of the WCR which can then
be used to calculate synthetic X-ray data. We review the steps behind
the construction of these models in Sec.~\ref{sec:xray_hydro}.

In short-period systems the WCR is often highly radiative. Due to the
difficulties involved in constructing hydrodynamical models under
these conditions, to date there have been few attempts to calculate
the X-ray emission from such systems. In Sec.~\ref{sec:igor} we review
recent work which takes a distinctly different approach which bypasses
problems which a traditional hydrodynamical simulation encounters in
this regime.

\subsection{Hydrodynamical models of the WCR} 
\label{sec:xray_hydro}

\begin{figure*}[ht]
\vspace{13cm}
  \begin{center}
\includegraphics{2.1_pittard_fig2a.ps}
\includegraphics{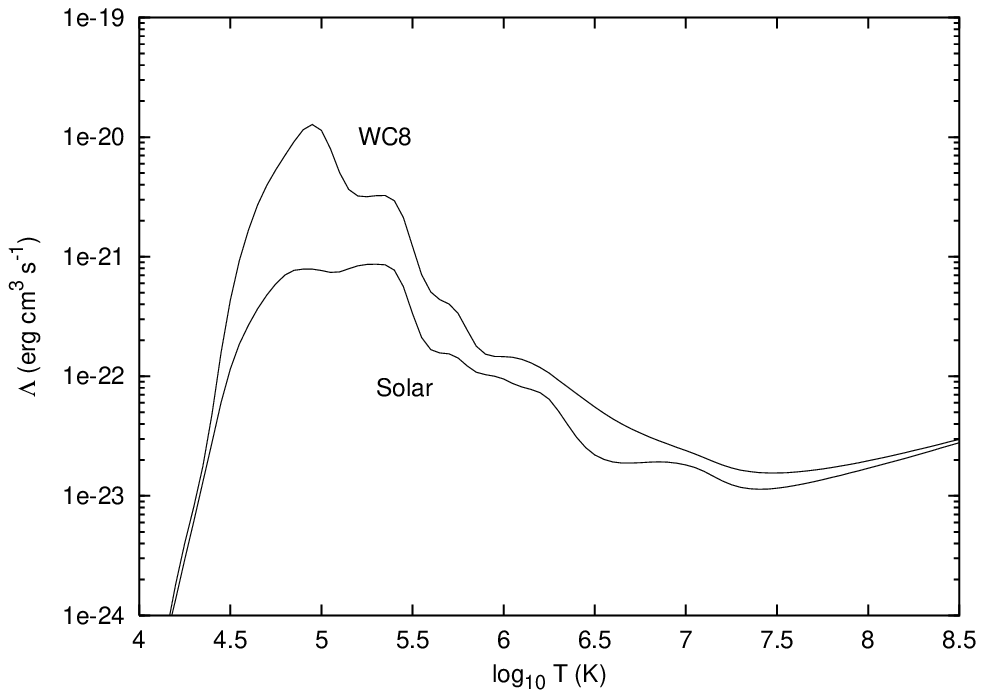}
\includegraphics{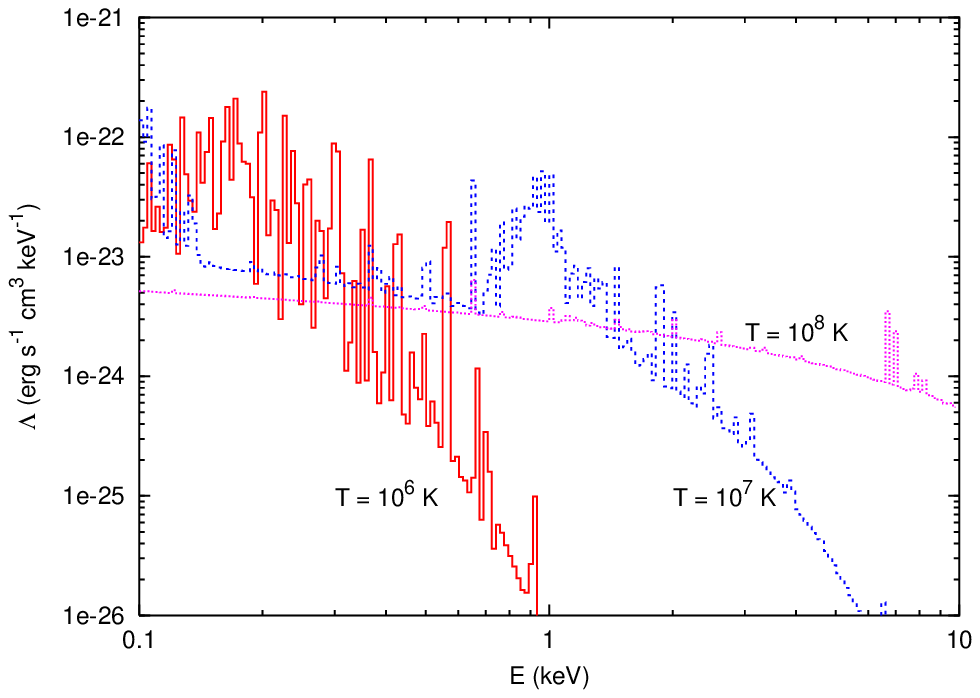}
\includegraphics{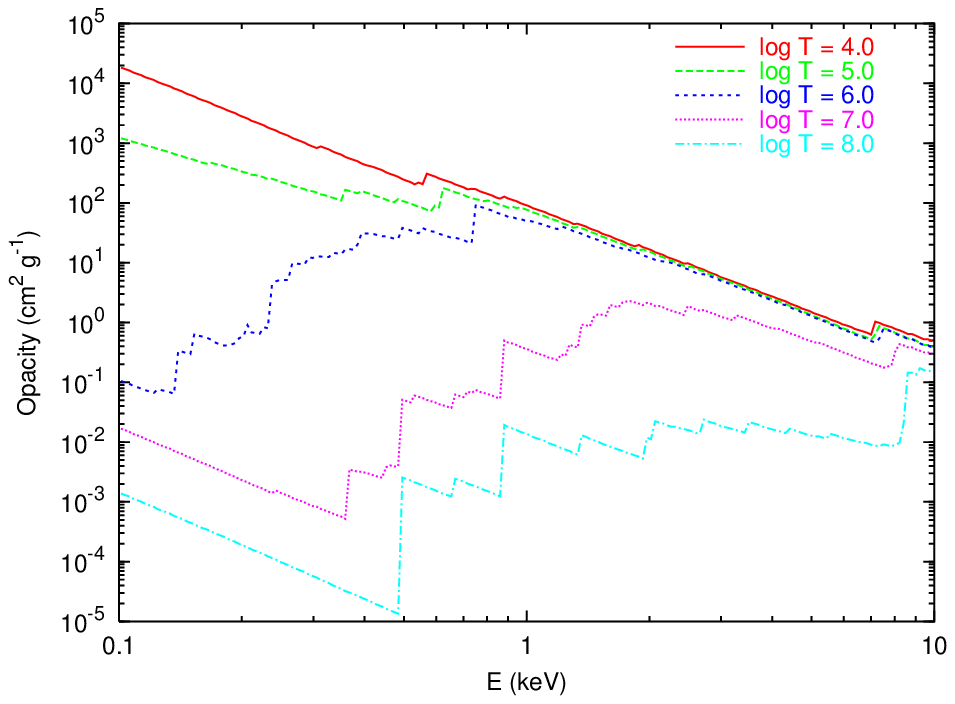}
    \caption{\small The main ingredients involved in the construction of
models of the X-ray
emission from colliding wind systems which are largely adiabatic. Top left:
a density plot from a hydrodynamical model of the wind-wind collision.
Top right: cooling curves for hot, thermal plasma. Bottom left: emissivity
data (for solar abundances). Bottom right: opacity data (for solar 
abundances).}
    \label{fig:xray_ingredients}
  \end{center}
\end{figure*}

Early numerical models were presented by (amongst others)
\citet{Lebedev:1988}, \citet{Luo:1990}, and \citet{Stevens:1992}.  The
main ingredients involved in these models are summarized in
Fig.~\ref{fig:xray_ingredients}. A hydrodynamical code is used to
model the dynamics of the WCR. In this example the
deflection of the WCR by the orbital motion of the stars is assumed to
be negligible, the WCR remains axis-symmetric, and a 2D $(r,z)$
calculation is performed. A small amount of artificial viscosity is
used in the calculations to prevent the development of the odd-even
decoupling (i.e. carbuncle)
instability at the apex of the collision zone. This numerical artifact
develops when shocks are locally aligned with the hydrodynamic grid
\citep{Walder:1994,Le Veque:1998}. In Fig.~\ref{fig:xray_ingredients},
$\eta = 0.1$, and the stellar winds were
assumed to have reached their terminal speeds. We also specified
$v_{\rm WR} = v_{\rm O}$, so the velocity shear along the contact
discontinuity is small and there is no noticeable development of the
Kelvin-Helmholtz instability (instabilities are discussed in greater
detail in Sec.~\ref{sec:instab}). In systems where the WCR is largely
adiabatic, it is not strictly necessary to compute the rate of energy
loss of the shocked gas through radiative cooling (since by definition
this is small).  However, it is often included in order that models
are self-consistent. The hot plasma is generally assumed to be
optically thin and in collisional ionization equilibrium, in which
case the cooling rate as a function of temperature may be calculated
using, for example, the MEKAL plasma emission code \citep{Mewe:1995}.

Cooling curves for solar abundances ($X$[H]$=0.705$, $Y$[He]$=0.205$,
$Z$[everything else]$=0.020$) and 
the abundances found for the WC8 star in $\gamma^{2}$~Vel ($X=0.0$, 
$Y=0.628$, $Z=0.372$) are shown in the top right panel of
Fig.~\ref{fig:xray_ingredients}. Line emission dominates the cooling
at temperatures below $10^{7} \K$. At higher temperatures cooling
occurs primarily through thermal bremsstrahlung ($\Lambda(T) \propto T^{1/2}$).
The enhanced metal abundance in the winds of WR stars leads to more
efficient cooling than is obtained for material of solar abundance.

To obtain synthetic X-ray data the hydrodynamical model is processed
through a ray-tracing code which solves the radiative transfer
equation along lines of sight through the hydrodynamical
grid. Emissivity and opacity data for solar abundance material are
shown in the bottom left and right panels of
Fig.~\ref{fig:xray_ingredients}, respectively.  The emissivity data
was again calculated using the MEKAL code, and consists of line
emission plus thermal bremsstrahlung continuum.  At $T = 10^{8} \K$
most elements are completely ionized and there is little line emission
in comparison to lower temperatures, while the bremsstrahlung emission
extends to higher energies. The opacity data in the bottom right panel
of Fig.~\ref{fig:xray_ingredients}\/ was calculated using the {\it
Cloudy} photoionization code\citep{Ferland:2001}. Distinct absorption
edges are clearly visible. As the temperature increases and elements
become more and more ionized the absorption cross-section decreases.

Once synthetic spectra and images have been calculated they can be
compared against actual X-ray data, and the properties of CWBs
determined. The most rigorous method involves the calculation of a
grid of synthetic X-ray data from a large number of hydrodynamical
models in which the stellar wind parameters are systematically
varied. An {\it XSPEC} table model of the synthetic X-ray data grid
can then be generated, and fitted against actual observational data.
Constraints on the stellar mass-loss rates and wind terminal speeds
have been obtained by this method for $\gamma^{2}$~Vel
\citep{Stevens:1996}, \wr140 \citep{Zhekov:2000}, and $\eta$~Carinae
\citep{Pittard:2002a}. In each of these works the X-ray derived
mass-loss rate for the star with the stronger wind is substantially
lower than the best estimate from radio observations at that
time. This is largely due to the fact that the X-ray derivation
appears to be insensitive to wind clumping (unlike radio-based
derivations of $\Mdot$). The X-ray based method also enjoys further
advantages. For instance, in the case of $\eta$~Carinae the high
energy X-ray's are much less affected by the complex circumstellar
environment than emission at longer wavelengths. An additional benefit
over other techniques is the ability to place constraints on the
properties of the star with the weaker wind.

\begin{figure}
  \begin{center}
\includegraphics[scale=0.35,angle=-90.0]{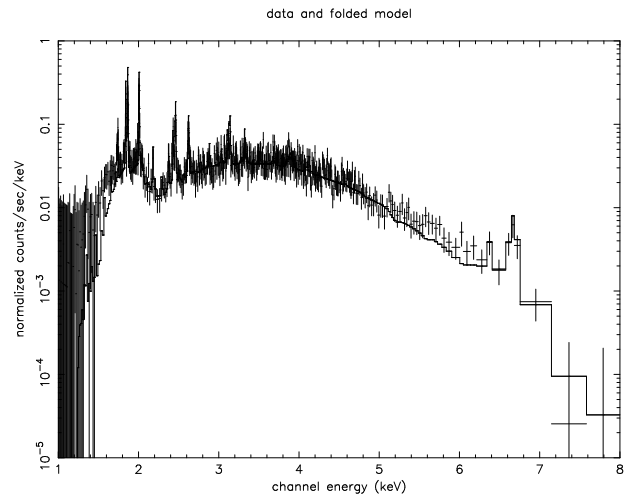}
    \caption{\small The best fit to a {\it Chandra} grating spectrum of
$\eta$~Carinae \citep[from][]{Pittard:2002a}. If we take the terminal speed 
of the primary wind as $v_{1} = 600 \;\kmps$, then the fit implies that 
$\Mdot_{1} = 2.5 \times 10^{-4} \; \Msolpyr$.}
    \label{fig:ec_specfit}
  \end{center}
\end{figure}

To illustrate this method at work, Fig.~\ref{fig:ec_specfit} shows a
fit to a {\it Chandra} grating spectrum of $\eta$~Carinae
\citep[reproduced from][]{Pittard:2002a}. $\eta$~Carinae is arguably
the most infamous massive star in our Galaxy, being best known as the
survivor of the greatest non-terminal stellar explosion ever
recorded. The ejecta have since formed the beautiful bipolar nebula
known as the Homunculus \cite[see, e.g.,][]{Morse:1998}. Study of
this nebula \cite[see, e.g.,][]{Smith:2003} plays a central role in the
quest for a deeper understanding of this object.  Unfortunately, it
seems that astronomers cannot have their cake and eat it, since the
nebula also acts as a screen which significantly obscures the central
star. For a long time $\eta$~Carinae was thought to be a single
massive star and was classified as an luminous blue variable (LBV), but there is now
overwhelming evidence for binarity and a strong wind-wind collision in
this system
\citep{Damineli:2000,Corcoran:2001b}.  Tantalizing evidence also
exists to relate the ``Great Eruption'' in 1843 with the close
approach of the stars during their passage through periastron. Whether
the tidal influence of the companion star was in part responsible for
triggering the enormous outburst is one of the most important
questions concerning $\eta$~Carinae today.

$\eta$~Carinae is in fact a highly unusual example of a CWB.  It may
be the only one known in which one of the components is an LBV star \citep[although HD~5980 may be another - see,
e.g.][]{Naze:2002}, and as such the low wind speed of the primary star
($v \sim 500 \kmps$) means that the hard X-ray emission (above $2
\keV$) is generated almost exclusively by shocked material from the
secondary wind. At the time of the {\it Chandra} observation (shown in
Fig.~\ref{fig:ec_specfit}) the attenuation of this hard emission by
the stellar winds is small and the line of sight into the system is
through the less dense wind of the companion star. This gives rise to
the unusual situation that the hard X-ray emission is not directly
sensitive to the mass-loss rate of the primary, but only indirectly
through the wind momentum ratio $\eta$. Fortunately, the duration of
the lightcurve minimum can be used to estimate $\eta$.  The best fit
to the X-ray data then gives $\Mdot \approx 10^{-5} \Msolpyr$ for the
secondary star and $v \approx 3000 \kmps$ for its wind. With
$\eta=0.2$ and assuming $v \sim 600 \kmps$ for the wind of the primary
star, the fit in Fig.~\ref{fig:ec_specfit} implies $\Mdot \approx 2.5
\times 10^{-4} \Msolpyr$ for the primary star. As noted earlier, this
value is substantially smaller than commonly inferred (in the
literature there is a large scatter in the derived values, though the
general view is that $\Mdot \sim 10^{-3} \Msolpyr$).  However, some
problems with a mass-loss rate as high as $10^{-3} \Msolpyr$ have been
identified: fits to {\it HST} data show absorption components which
are too strong and electron-scattering wings which were overestimated
\citep{Hillier:2001}. Likewise, the X-ray derived results face their
own set of problems. For instance, the derived wind parameters of the
secondary star suggest that it is either a very massive O-star or
perhaps a WR star, but no evidence for a companion has been found to
date in optical spectra.  It will be interesting, therefore, to see if
the results in \citet{Pittard:2002a} are consistent with those from
further X-ray observations.  Analysis of additional {\it Chandra}
spectra taken through periastron passage is ongoing, and we hope to
report results soon.

Since the launch of {\it Chandra} and {\it XMM-Newton}, it has also become
possible to measure the shift and width of individual X-ray lines from
CWBs. The high spectral resolution offered by the
gratings on these telescopes allows access to a wide range of important
diagnostic data which can be used to place additional constraints on
the wind-wind collision. The theoretical study of line profiles in
CWBs is in its infancy, although some unexpected results have already 
been obtained. We refer the interested reader to 
Henley et al. (these proceedings).

\subsection{Modeling X-ray emission from highly radiative systems}
\label{sec:igor}
It is impossible to use the techniques described in the previous
section when the shocked gas is highly radiative because in order to
calculate the emission the cooling length must be spatially resolved.
Hydrodynamical models (including those with adaptive-mesh-refinement)
impose a limit to the spatial resolution which can be much larger than
the cooling length of the gas in such systems. Another problem is that
the rapid cooling of hot gas creates cold dense sheets which are
unstable to a variety of instabilities \citep[see,
e.g.,][]{Stevens:1992}, and which introduce a large time dependence
into the computed emission.

These short-comings have been recently addressed by
\citet{Antokhin:2004}.  Their solution is to separate the calculation
of the small-scale (local) cooling structure of the post-shock gas
from the global structure of the wind ram-pressure balance surface, and to
assume that these structures are time-independent. The steady-state
shape of the contact discontinuity in a CWB has been
comprehensively discussed in the literature \citep[see references
in][]{Antokhin:2004}, though in these works the winds were assumed to
have reached their terminal speeds before colliding.  However, as the
post-shock gas becomes increasingly radiative with decreasing stellar
separation, such models are most applicable to short-period systems
where the wind speeds will often have not reached their terminal
values at the interaction front. To account for this,
\citet{Antokhin:2004} assumed ``beta''\footnote{A beta velocity law has the form:
$$v(r) = v_{\infty} (1 - R/r)^{\beta}$$ where R is the radius of the
star. $\beta=1$ is often used in analytical works, $\beta=0.8$ is
fairly representative of O-stars, while WR stars have more slowly
accelerating winds (which means higher values of beta, though such a
formulation is probably very poor for WR stars).} velocity laws in
their model.


In the highly radiative limit, the timescale for shock heated material
to cool is small compared to the flow time for material to advect
along a substantial arc of the contact discontinuity, and the
post-shock gas can be assumed to occupy a layer which is geometrically
thin with respect to other scales in the system (e.g., the binary
separation, and the radius of curvature of the contact
discontinuity). The expansion of the shocked gas as it flows out of
the system normally couples the internal evolution of the post-shock
gas to the global wind-wind structure, since adiabatic cooling can
then take place. However, with the above assumptions one can adopt an
``on-the-spot'' treatment of the radiative emission at each point
along the interaction front, and the post-shock structure can be
simplified to planar geometry.

\begin{figure}
  \begin{center}
\includegraphics[scale=0.35]{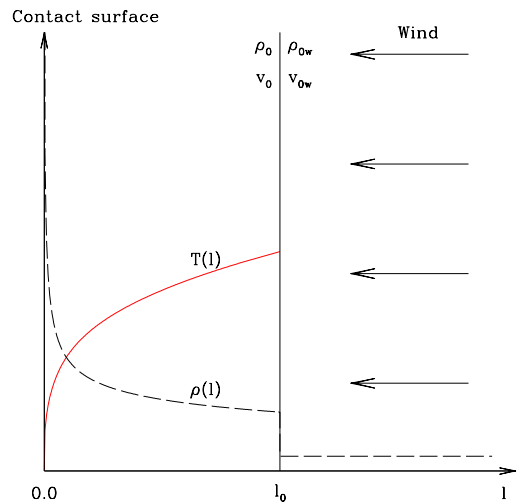}
    \caption{\small A schematic of the post-shock density and temperature
structure in a steady, planar cooling layer (reproduced from 
\citet{Antokhin:2004}, with permission). The shock front occurs at $l_{0}$.}
    \label{fig:ps_structure}
  \end{center}
\end{figure}

To obtain the structure of the cooling layer, \citet{Antokhin:2004}
adopted an isobaric approximation. Assuming that the energy loss in the 
shocked gas is solely from radiative cooling, the rate of temperature 
decrease in the post-shock gas is given by,
\begin{equation}
\label{eq:ps_cool}
\frac{5k}{2 \mu m_{\rm p}}\rho v \frac{dT}{dl} = -n_{\rm e}n_{\rm H}\Lambda(T),
\end{equation} 
where $l$ is a distance from the shock front, and $\Lambda(T)$ is the cooling
function. Equation~\ref{eq:ps_cool}
also gives the width of the cooling layer, and a schematic of the post-shock
structure is shown in Fig.~\ref{fig:ps_structure}.

\begin{figure}
  \begin{center}
\includegraphics[scale=0.35]{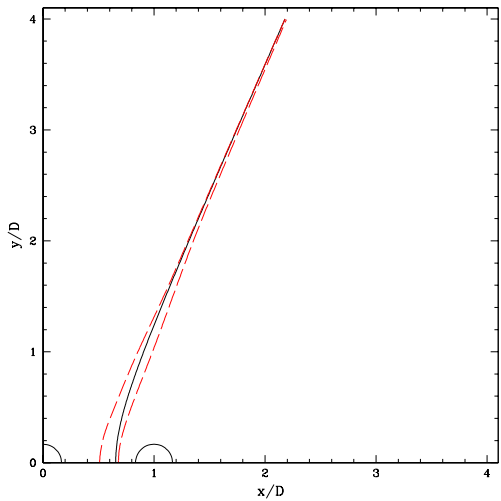}
    \caption{\small The structure of the wind-wind interaction from the
model of \citet{Antokhin:2004} (reproduced with permission). The solid line
marks the contact discontinuity, while the dashed lines indicate the 
shock fronts.}
    \label{fig:global_structure}
  \end{center}
\end{figure}

An example of the global structure of the WCR,
together with the width of the post-shock cooling layer either side of
the contact discontinuity, is shown in
Fig.~\ref{fig:global_structure}. An initial application of this
method has been made to an {\it XMM-Newton} observation of the
short-period binary HD~159176 \citep{DeBecker:2004}, with good success
achieved in fitting the spectral shape. However, in this work it was not
possible to obtain a good fit which was simultaneously in agreement
with the predicted luminosity from the model.

This method shows great promise and it will clearly be useful to
develop it further. Future work on this model could include extension
to three dimentions (which would allow deflection of the interaction
region into a spiral form as a result of orbital motion of the stars),
and the dynamical role of the stellar radiation fields \citep[see,
e.g.,][and further discussion in Sec.~\ref{sec:other}]
{Stevens:1994,Gayley:1997}. Further tests against a variety of
short-period CWBs also need to be made, and in this respect {\it
XMM-Newton} data from the archetype system V444~Cygni scheduled for
May-June 2004 are ideally suited. It should also prove illuminating to
compare the results from this method and those from hydrodynamical
models in the transition region where $\chi \sim 1$.

\section{Radio emission from early-type stars}
\label{sec:radio}
Radio observations of early-type stars reveal that they can emit both
thermal and nonthermal radiation. The former is readily explained as
arising from free-free emission within the strong stellar wind
\citep{Wright:1975}. 
The winds of early-type stars are partially optically thick at
radio wavelengths, and the optical depth at radius $r$ is given by
\begin{equation}
\label{eq:opt_depth}
\tau_{\nu}(r) \approx \int^{\infty}_{r} 10^{-4} \left(\frac{Z^{2}\gamma_{e}}
{\mu_{i}^{2}}\right) \left(\frac{\Mdot}{v_{\infty}}\right)^{2} 
\frac{g_{\nu}}{m_{\rm H}^{2} T^{3/2} \nu^{2}} \frac{1}{r^{4}} dr,
\end{equation}
where $Z$ is the rms charge of the atoms ($Z=1$ for
singly ionized gas), $\gamma_{e}$ is the number of free electrons per
ion, $\mu_{i}$ is the mean atomic mass of the ions in units of $m_{\rm
H}$, $g_{\nu}$ is the gaunt factor and $T$ is the temperature of the
wind \citep[see, e.g.,][]{Lamers:1999}. The high mass-loss rate of
massive stars means that the photospheric radius at radio wavelengths
is always much larger than the radius of the underlying star.  For an
isothermal O-star wind with $T=10,000\K$, $\Mdot = 10^{-6}\Msolpyr$,
$v_{\infty}=2000\kmps$, and consisting mainly of ionized hydrogen, the
$\tau_{\nu} = 1$ surface occurs at radii of 1350, 1900 and $4200\Rsol$
at radio wavelengths of 3.6, 6 and $20\cm$ respectively. For WR stars,
these radii are even larger. Since the
wind speed is assumed to have reached its terminal value at these
distances, radio observations have proved to be a useful tool for
measurements of mass-loss rates (although as discussed in 
Sec.~\ref{sec:xray_hydro} these suffer from uncertainties due to the
clumpy nature of such winds).

The nonthermal emission, on the other hand, is attributed to
synchrotron radiation from relativistic electrons moving in a magnetic
field.  The consensus is that the relativistic electrons in the winds
of massive stars are created through diffusive shock acceleration. In
this process the electrons gain energy through the first-order Fermi
mechanism as they cross and re-cross the shock until they are
eventually scattered downstream \citep[see,
e.g.,][]{Bell:1978}. Electrons with Lorentz factor $\gamma \ltsimm
10^{5}$ are frozen into the post-shock flow \citep{White:1985}. The
resulting energy distribution of the relativistic electrons has a
power law form, i.e. the number of relativistic electrons per unit
volume with energy between $\gamma$ and $\gamma + d\gamma$, $N(\gamma)
\propto \gamma^{-n}$. The energy index, $n$, depends on the shock
compression ratio, $\chi_{c}$, through \citep{Bell:1978}
\begin{equation}
\label{eq:chi_c}
n = \frac{\chi_{c}+2}{\chi_{c}-1}.
\end{equation}
In the case where the shocked plasma radiates efficiently, the shock is
quasi-isothermal, and the compression ratio can become very large,
leading to $n \sim 1$. If instead the shock is adiabatic, then $\chi_{c}=4$
(for a strong shock), and $n=2$.

As already noted, hydrodynamical shocks are a feature of the winds of
both single massive stars and their binary counterparts. In the case
of single stars, the intrinsic instability of the radiatively driven
wind leads to the formation of shocks which are distributed throughout
the wind \citep[see][and references therein]{Runacres:2002}. While in
principle these can accelerate electrons to high energies, the intense
stellar radiation field causes significant energy losses through
inverse Compton (IC) scattering and electrons accelerated in the inner
regions of the stellar wind cannot survive out to large radii. Those
electrons which are responsible for the nonthermal radio emission must
therefore be accelerated at large distances from the star
\citep{Chen:1994}. The large photospheric radii of massive stars at
radio wavelengths imposes a similar constraint, since for nonthermal
emission to be observed at radio wavelengths, the relativistic electrons 
must exist outside the radio photospheric radius of the stellar wind.

In binary systems the strong shocks formed by the collision of the two
winds is a natural place for diffusive shock acceleration of
nonthermal electrons. Another possibility
concerns the current sheets which result from the collision of
magnetized stellar winds \citep{Jardine:1996}. The compression of
magnetic field lines in the current sheets create enhanced local field
strengths, and the resulting electric fields may accelerate particles
to high energies. Irrespective of which is the dominant acceleration
mechanism, wide binary systems have the significant advantages that
the wind collision may exist well outside the radio photosphere of the
stars, and that the stellar radiation energy density is substantially
diluted in the vicinity of the WCR. This is reflected in strong
observational support for a CWB interpretation for all synchrotron
emission from WR stars \citep{Dougherty:2000b}, though the case is not
so clear for O-stars \citep[see, e.g.,][]{Rauw:2002}. 

However, when synchrotron emission is not detected, that does not mean
it is not being emitted. Nonthermal radio emission generated in the
WCR of a short period binary may be buried so deep under the radio
photosphere that the heavy opacity prevents its observation. For the
nonthermal emission to be highly visible we require that the binary is
wide enough that the apex of the WCR is far enough away from the stars
that it exists outside of their radio photospheres. This separation
implies a value of the orbital period of order of a month. Therefore,
there is a selection effect that all the CWBs with observed
synchrotron are wide binaries.

\subsection{Modeling the radio emission from CWBs}
Most analysis of the radio spectra from colliding wind binaries
has hitherto been based on a highly simplified model consisting of a
thermal source plus a point-like source of nonthermal emission
attenuated by free-free absorption \citep[e.g.,][]{Chapman:1999,
Monnier:2002}. In this model, at frequency $\nu$, the 
observed flux ($S_\nu^{\rm
obs}$) is related to the thermal flux ($S_\nu^{\rm th}$), the
synchrotron emission arising from the WCR 
($S_\nu^{\rm syn}$), and the free-free opacity ( $\tau_\nu^{\rm ff}$)
of the circum-system stellar wind envelope by
\begin{equation}
\label{eq:radio_simple}
S_{\nu}^{\rm obs} = S_{\nu}^{\rm th} + S_{\nu}^{\rm syn} {\rm e}^{-\tau_{\nu}^{\rm ff}}.
\end{equation} 
From simultaneous measurements in a number of frequency bands it is
possible to solve Eq.~\ref{eq:radio_simple} to determine the
relative contribution of thermal and nonthermal emission to the total flux.
On the other hand, both $S_{\nu}^{\rm syn}$ and $\tau_{\nu}^{\rm ff}$ may 
display significant time variability. The former is expected if the
stellar separation varied as in an eccentric binary, while the latter
is in general anticipated since the optical depth to the synchrotron emitting
region will change as the line of sight sweeps around the system.

While Eq.~\ref{eq:radio_simple} may at times be adequate, 
a single-valued free-free opacity determined along
the line-of-sight to a point-like synchrotron emission region is a
clear over-simplification: observations reveal an extended region of
synchrotron emission from both WR~146
\citep{Dougherty:1996,Dougherty:2000a} and WR~147
\citep{Moran:1989,Churchwell:1992,Williams:1997} - see also
Sec.~\ref{sec:wr147}. Furthermore, use of such over-simplified models
has often resulted in difficulties in fitting observational
data. Nowhere are these problems better illustrated than for the
wonderful dataset which exists for WR~140, the archetype wide CWB
\citep[see, e.g.,][]{Williams:1990, White:1995}.
We note that the effect of binarity on the {\em thermal} radio emission from 
massive stars has also been investigated \citep{Stevens:1995}, though
this work did not examine nonthermal emission.

Complex magnetohydrodynamic models which include the transport of
nonthermal electrons have been computed to predict the synchrotron
emission from other astrophysical objects, such as radio jets
\citep{Tregillis:2004}. However, less attention has been focused on
models for CWBs, and the resulting models are less well developed.
In fact, CWBs are nice laboratories for particle acceleration because
the fundamental flow parameters of the winds are well known.

As a first step towards the construction of more realistic models,
\citet{Dougherty:2003} obtained a simple representation of the spatial
distribution of the free-free and nonthermal emission from CWB
systems through the use of a hydrodynamical model of the WCR.  In this
section we note the main results of this work.

\begin{figure}[t]
\begin{center}
\includegraphics[scale=0.6]{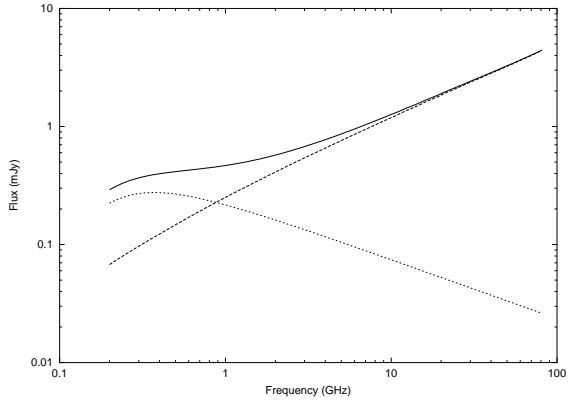}
\caption{\small Spectra from the standard model with $0^\circ$ inclination
  - free-free flux (dashed), magneto-bremsstrahlung flux (dotted), and
  total flux (solid).  IC cooling, coulombic cooling, the
  Razin effect and synchrotron self absorbtion (SSA) are not included in this calculation. 
}
\label{fig:stan_spec1}
  \end{center}
\end{figure}

The process of calculating the theoretical radio emission from a 
hydrodynamical model of a CWB is as follows.  
First, the temperature and density values on the
hydrodynamic grid are used to calculate the free-free emission and
absorption coefficients from each grid cell. The synchrotron emission
and synchrotron self-absorption from each cell within the WCR is
calculated assuming that the distribution of relativistic electrons in
each cell can be specified by $N(\gamma) d\gamma = C \gamma^{-p} d\gamma$,
where $p=2$. The standard
assumption that
$C$ is proportional to the thermal energy density is made \citep[c.f.][]
{Chevalier:1982,Mioduszewski:2001}, and it is assumed that $p$ is
spatially invariant over the volume of the WCR. 
The magnetic field in the WCR is assumed to be highly tangled, and the 
synchrotron emissivity ($P_{\nu} \propto \nu^{-(p-1)/2}$) is
assumed to be valid over all frequencies, $\nu$. A radiative transfer
calculation is then computed to obtain the flux and intensity
distribution at a specified frequency. The interested reader is referred
to \citet{Dougherty:2003} for further details.

\begin{table}[t]
  \begin{center}
  \caption{Parameters for the ``standard'' CWB model in 
\citet{Dougherty:2003}.}\medskip
  \label{tab:standard_radiocwb}
  \begin{tabular}{ll}
    Parameter & Value \\
    \hline\hline
    $\Mdot_{\rm WR}\;(\Msolpyr)$ & $2 \times 10^{-5}$ \\
    $\Mdot_{\rm O}\;(\Msolpyr)$ & $2 \times 10^{-6}$ \\
    $v_{\infty,{\rm WR}}\;(\kmps)$ & $2000$ \\
    $v_{\infty,{\rm O}}\;(\kmps)$ & $2000$ \\
    $D\;(\cm)$ & $2 \times 10^{15}$ \\
    \hline
  \end{tabular}
  \end{center}
\end{table}

We shall first discuss the results obtained from a ``standard'' CWB
model with the parameters noted in Table~\ref{tab:standard_radiocwb}.
These wind values are typical of WR and OB stars, and with the adopted
binary separation, $D$, the WCR is largely
adiabatic.  At a frequency $\nu = 5\;{\rm GHz}$, $D$ is approximately
$10\times$ the radius of the $\tau_{\rm ff}=1$ surface of the WR
wind. For inclination angles $\sim$$0^\circ $, the lines of sight to
the WCR are then optically thin, permitting
investigation of the emission from the wind-collision region in the
absence of strong free-free absorption from the circum-binary stellar
wind envelope. Solar abundances for the O-star and WC-type abundances
for the WR-star (mass fractions $X=0,Y=0.75,Z=0.25$) are adopted in
the model. Temperatures of $10,000$~K and an ionization structure of
${\rm H^{+}}$, ${\rm He^{+}}$ and ${\rm CNO^{2+}}$ are assumed for the
unshocked stellar winds, and the system is assumed to be positioned at
a distance of 1.0~kpc. Finally, it is assumed that the magnetic energy
density, $U_{B}$, the relativistic electron energy density, $U_{\rm
rel}$, and the thermal energy density, $U_{\rm th}$, are related by
\begin{equation}
\label{eq:b_en_dens}
U_{\rm B} = \frac{B^{2}}{8 \pi} = \zeta_{\rm B} U_{\rm th},
\end{equation}
and
\begin{equation}
\label{eq:nt_en_dens}
U_{\rm rel} = \int n(\gamma) \gamma mc^{2} d\gamma = \zeta_{\rm rel}
U_{\rm th},
\end{equation}
where $\zeta_{\rm B}$ and $\zeta_{\rm rel}$ are constants, and $U_{\rm
th}={P\over{\Gamma-1}}$ where $P$ is the gas pressure and $\Gamma$ is
the adiabatic index (assumed to be $5/3$, as for an ideal gas). The
common assumption of equipartition corresponds to $\zeta_{\rm B} =
\zeta_{\rm rel} = \zeta$.  The nonthermal
electrons lose energy due to synchrotron emission and IC 
scattering in the same ratio as the magnetic field energy
density to the photon energy density,
\begin{equation}
\frac{P_{\rm sync}}{P_{\rm compt}} = \frac{U_{\rm B}}{U_{\rm ph}}.
\end{equation}
For all reasonable values of $\zeta_{\rm
B}$, $P_{\rm sync} < P_{\rm compt}$, and IC cooling
is the dominant energy loss mechanism. Synchrotron
cooling is therefore ignored.

\begin{figure}[!h]
\begin{center}
\includegraphics[scale=0.45]{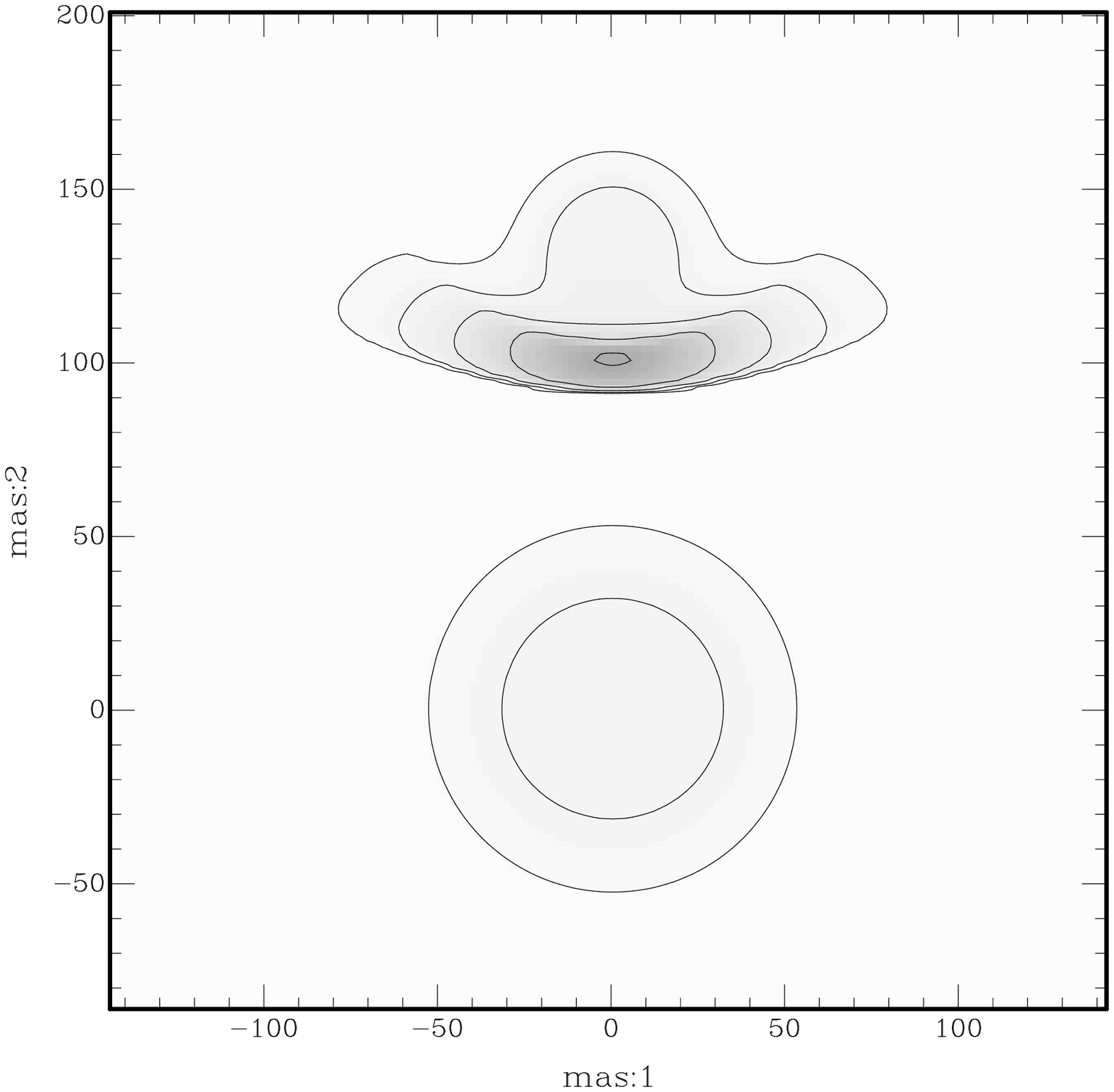}
\includegraphics[scale=0.45]{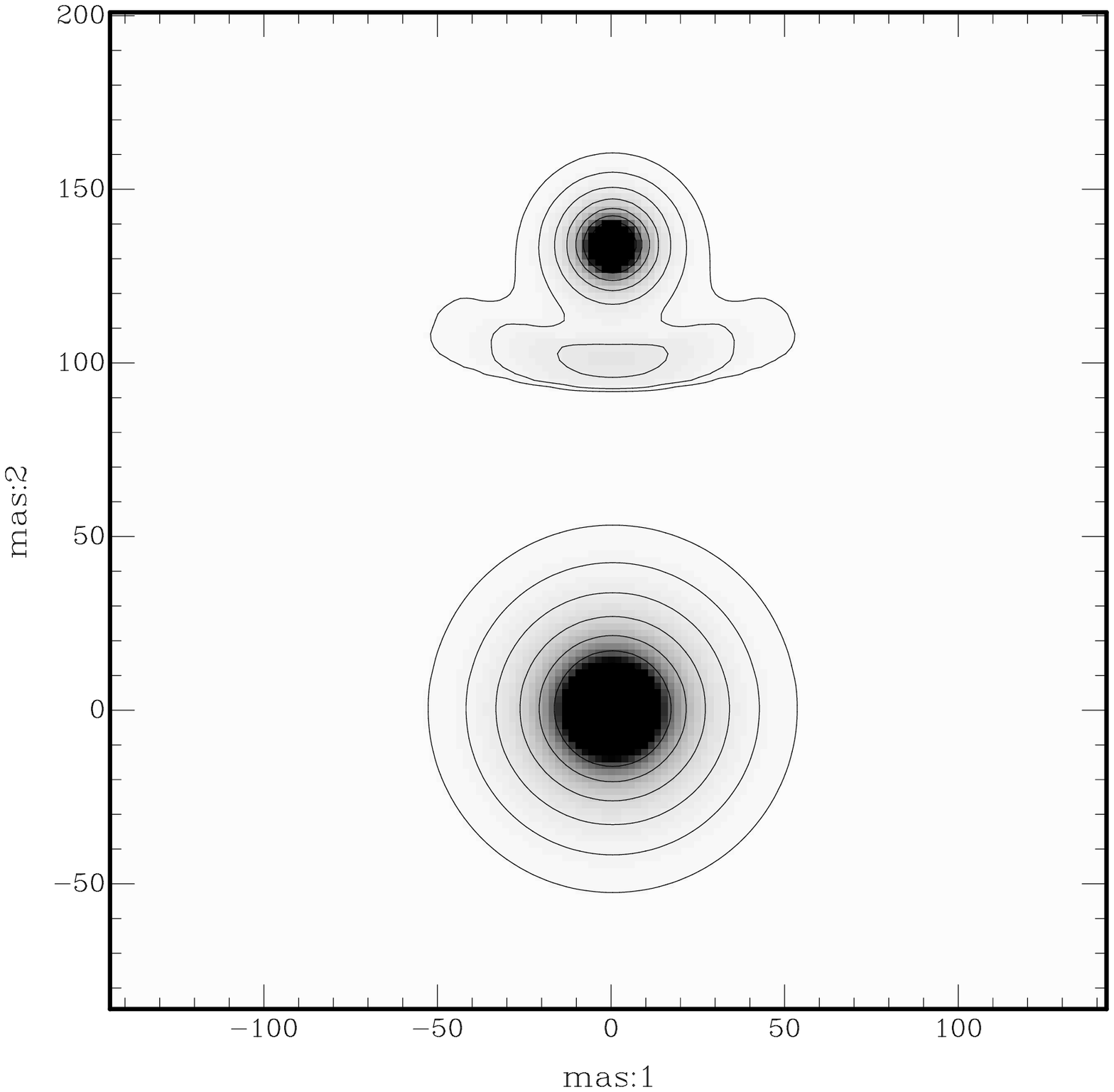}
\caption{\small Intensity distributions from the standard model at $0^\circ$
  inclination at 1.6~GHz (top) and 22~GHz (bottom). Neither IC
  cooling, coulombic cooling, the Razin effect or SSA, are included in this
  calculation. The stellar winds and the WCR are clearly visible.} 
\label{fig:stan_image1}
\end{center}
\end{figure}

\begin{table}[t]
  \begin{center}
  \caption{Values from the ``standard'' CWB model in 
\citet{Dougherty:2003}.}\medskip
  \label{tab:standard_params}
  \begin{tabular}{ll}
    Parameter & Value \\
    \hline\hline
    $n_{\rm max}\;(\pcm3)$ & $4\times 10^{5}$ \\
    $T_{\rm max}\;(\K)$ & $2\times 10^{8}$ \\
    $B_{\rm max}\;(\mG)$ & 6 \\
    \hline
  \end{tabular}
  \end{center}
\end{table}

The maximum post-shock
density, temperature, and B-field occur at the stagnation point
between the stars and are listed in Table~\ref{tab:standard_params}.
The synchrotron emission arising from a
single relativistic electron with Lorentz factor $\gamma$ is
\citep[c.f.][]{Rybicki:1979}
\begin{equation}
\label{eq:sync_single}
P(\nu) = \frac{\sqrt{3}q^{3}B{\rm sin}\alpha}{\rm m_{e}c^{2}}F(\nu/\nu_c),
\end{equation}
where $q$ is the electron charge, $B$ is the magnetic field
strength, $\alpha$ is the pitch angle of the particle relative to the
direction of the B-field, ${\rm m_{e}}$ is the mass of the electron, 
$c$ is the speed of light, and
$F(\nu/\nu_c)$ is a dimensionless function describing the total power
spectrum of the synchrotron emission, with $\nu_c$ the frequency where
the spectrum cuts off, given by
\begin{equation}
\label{eq:nu_c}
\nu_{\rm c}=\frac{3\gamma^2 q B {\rm sin}\alpha}{\rm 4 \pi m_{e} c}. 
\end{equation}
Values for $F(\nu/\nu_c)$ are tabulated in
\citet{Ginzburg:1965} and we assume that
${\rm sin}~\alpha = 1$. At the stagnation point $\nu_{\rm c}=0.25$~GHz
(for $\gamma=10^{2}$) and 25~GHz (for $\gamma=10^{3}$).

The spectra and intensity distributions obtained from the standard
model with $\zeta = 10^{-4}$ are shown in Fig.~\ref{fig:stan_spec1}
and Fig.~\ref{fig:stan_image1} respectively where, for simplicity,
only free-free absorption is included.  Magneto-bremsstrahlung
emission dominates the total flux below 1~GHz.  The turnover seen near
400~MHz is due to free-free absorption from the unshocked stellar
winds. Above 400~MHz, the magneto-bremsstrahlung spectrum is a
power-law with a spectral index ($S_\nu \propto \nu^\alpha$) $\alpha =
-0.5$, as expected for a $p=2$ energy distribution.  The thermal
spectrum spectral index of $+0.6$ is also as expected, though a slight
departure from a power-law exists at low frequencies due to the finite
size of the hydrodynamical grid. The intensity distributions shown in
Fig.~\ref{fig:stan_image1} reveal that at 1.6~GHz the WCR is
brighter than the stellar winds, while the opposite is true at 22~GHz.

\citet{Dougherty:2003} also explored how the radio flux varied with
stellar separation, and we repeat the main conclusions here.  We shall
first consider how varying the stellar separation affects the
free-free opacity to a specific point in the WCR.  For a stellar wind
with an $r^{-2}$ radial density distribution the line-of-sight opacity
$\tau$, at frequency $\nu$ through the wind is $\tau\propto
\xi^{-3}\nu^{-2.1}$, where $\xi$ is the impact parameter. Since $\xi
\propto D$ for a given inclination, the turnover frequency for a
constant opacity value is $\nu\propto D^{-10/7}$.
\citet{Dougherty:2003} showed that this relationship is in excellent
agreement with the data shown in their Fig.~5.

Now let us consider how the {\em intrinsic} synchrotron emission
depends on the stellar separation. In the absence of synchrotron self
absorbtion (SSA) or the Razin 
effect, the synchrotron
luminosity increases as the separation decreases, due to the increased
thermal energy density in the collision region. In this scenario, the
intrinsic synchrotron emission per unit volume $P(\nu) \propto
\zeta^{3/4}n^{3/4}\nu^{-1/2}$ for an electron power-law index
$p=2$. Since the post-shock density $\propto D^{-2}$ and the volume of
the emitting region scales as $D^{3}$, the total synchrotron emission
from the entire wind collision volume scales as $D^{-1/2}\nu^{-1/2}$,
as noted by \citet{Dougherty:2003} and displayed in their Fig.~6.

\begin{figure}[!h]
\begin{center}
\includegraphics[scale=0.66]{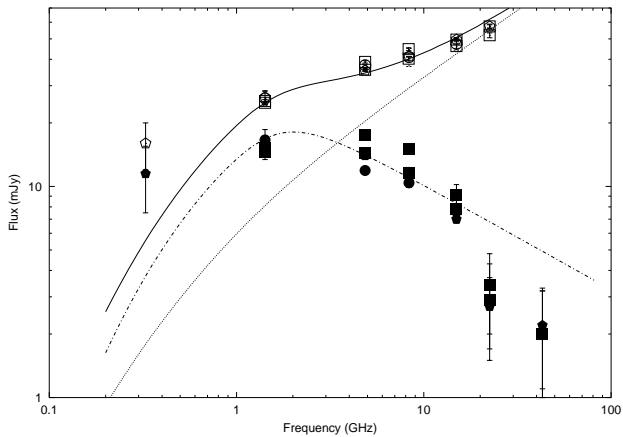}
\caption{\small The radio spectrum of WR 147. The deduced synchrotron
emission is represented by solid data points, while the total flux is
represented by the hollow data points. Estimates from different
authors are as follows: solid squares \citep[two separate estimates from 
the same observational data,][]{Skinner:1999}, solid pentagons
\citep{SetiaGunawan:2001}, open circles \citep[][]{Churchwell:1992,
Contreras:1999}. The lines are the total (solid), thermal (dotted) 
and synchrotron (dot-dashed) spectra of the $i = 0^{\circ}$
model with $\zeta =7.03\times 10^{-3}$, where SSA, Razin and free-free
absorption are included in the radiative transfer calculations. The
model winds were assumed to be clumpy with a volume-filling factor
$f=0.134$. Models at $+30^\circ $ and $-30^\circ $ are also consistent
with the data.}
\label{fig:wr147_spec}
\end{center}
\end{figure}

\subsubsection{Modeling the radio emission from WR~147}
\label{sec:wr147}
WR~147 is notable for being among the brightest WR stars at radio
frequencies, and for being one of two systems in which the thermal and
synchrotron emission are observed to arise from two spatially resolved
regions \citep[see, e.g.,][and references therein]{Williams:1997}.
Furthermore, it is one of a handful of WR+OB binary systems
where the two stars are resolved into a visual pair \citep{Niemela:1998,
Williams:1997} with a projected stellar separation given by 
$D\cos~i=0.635\pm0.020$\farcs.  At the estimated distance of $\sim$0.65~kpc
\citep{Churchwell:1992,Morris:2000} this corresponds to a
separation $D\sim415/\cos~i$~AU, and the relationship between $D$ and $i$
represents an important constraint for any models of the system. As the
inclination angle is unknown, \citet{Dougherty:2003} investigated 
models for several different values of $i$, which in turn requires 
different values of the physical
separation, $D$, to maintain the observed angular separation.

The radio spectrum of WR 147 is perhaps the best observed of all
massive binary systems, with radiometry extending from 353~MHz to
42.9~GHz. However, differences in excess of 50\% exist in the derived
synchrotron flux at some frequencies (cf. Fig.~\ref{fig:wr147_spec}),
giving rise to uncertainty in the position of the low frequency
turnover \citep[see][for further details]{Dougherty:2003}.
These differences form the fundamental reason for conflicting
conclusions about the nature of the underlying electron energy
spectrum in the current literature \citep{Skinner:1999,SetiaGunawan:2001}.

Full details of the modeling of the radio spectrum of WR~147 can be
found in \citet{Dougherty:2003}.  Values for
the mass-loss rates of the stars, and the terminal velocities of the
winds are noted in Table~\ref{tab:wr147_params}. These were taken from
the literature, as were details of the wind compositions.  The
temperature for both stellar winds was assumed to be 10~kK, and the
dominant ionization states were taken to be H$^{+}$, He$^{+}$ and
CNO$^{2+}$.  The thermal flux is only weakly dependent on the assumed
wind temperature (through the Gaunt factor), if the ionization state
is fixed.

\begin{table}[t]
  \begin{center}
  \caption{Parameters for the WR~147 model in 
\citet{Dougherty:2003}.}\medskip
  \label{tab:wr147_params}
  \begin{tabular}{ll}
    Parameter & Value \\
    \hline\hline
    $\Mdot_{\rm WN8}\;(\Msolpyr)$ & $2 \times 10^{-5}$ \\
    $\Mdot_{\rm OB}\;(\Msolpyr)$ & $3.8 \times 10^{-7}$ \\
    $v_{\infty,{\rm WN8}}\;(\kmps)$ & $950$ \\
    $v_{\infty,{\rm OB}}\;(\kmps)$ & $1000$ \\
    \hline
  \end{tabular}
  \end{center}
\end{table}

Models with $i=0$, 30 and $60^{\circ}$ were computed, and values of
$\zeta$ were estimated for each model by matching the mean of the
``observed'' synchrotron 4.86~GHz fluxes, which was taken to be
$14.1\pm0.3$~mJy. The spectrum between 0.2 and 80~GHz was then
calculated assuming this constant value of $\zeta$. 
To fit the thermal flux with the chosen value of $\hbox{${\dot M}_{\rm WN8}$ }$
the winds were required to be clumped.

In Fig.~\ref{fig:wr147_spec} the resulting total, synchrotron and
thermal spectra from one of the models ($i=0$, and including SSA, the
Razin effect and free-free absorption) is shown against the observed
spectra. Though this is not the best fit to the data in a formal
sense, the model fits the data reasonably well given the
approximations and assumptions which it contains. The shortcomings are
that the total flux in the 5 to 8~GHz range is underestimated, and the
synchrotron emission does not turn down somewhere around 10 to 20~GHz
as suggested by the observed data. The shortfall in the 5-8~GHz total
flux could be eliminated by increasing the thermal emission by a few
mJy. This could be achieved by simply increasing $\hbox{${\dot M}$
}/v_\infty$ in the stellar winds. While this would require that the
synchrotron emission at frequencies higher than $\sim$ 10~GHz be lower
than in the current models, this may well be the case if IC cooling is
taken into account, as explained below.

The fit shown in Fig.~\ref{fig:wr147_spec} has $\zeta \approx 7 \times
10^{-3}$, which implies that $B_{\rm max} = 4\mG$. If the tangled
field in the WCR has its origin in the O-star photosphere we deduce
that the B-field at the surface of the O-star is $B_{*} \sim 100$~G. This
is comparable with the non-detection of the Zeeman effect for the
majority of OB stars which puts an upper limit $\sim 100$~G on $B_{*}$
\citep{Mathys:1999}.  It is also worth noting that it is assumed that the
nonthermal {\em electron} energy density is in equipartition with the
magnetic and thermal energy densities. Since the energy density of
nonthermal protons, $U_{\rm p}$, is likely to be $50\times$ higher 
than that of the electrons, the deduced value of 
$\zeta \approx 7 \times 10^{-3}$ means that $U_{\rm p}$ is
uncomfortably close to unity. If the shocks in CWBs really are
this efficient at accelerating particles, the thermal properties of the 
shock-heated, X-ray emitting gas will be affected. It is interesting to
note that X-ray spectra from CWBs has to date
always been modeled and interpreted assuming that the shocks place 
an insignificant fraction of their energy in relativistic particles,
so this finding has implications for the analysis of X-ray data.

\begin{figure*}[t]
\begin{center}
\includegraphics[scale=0.55]{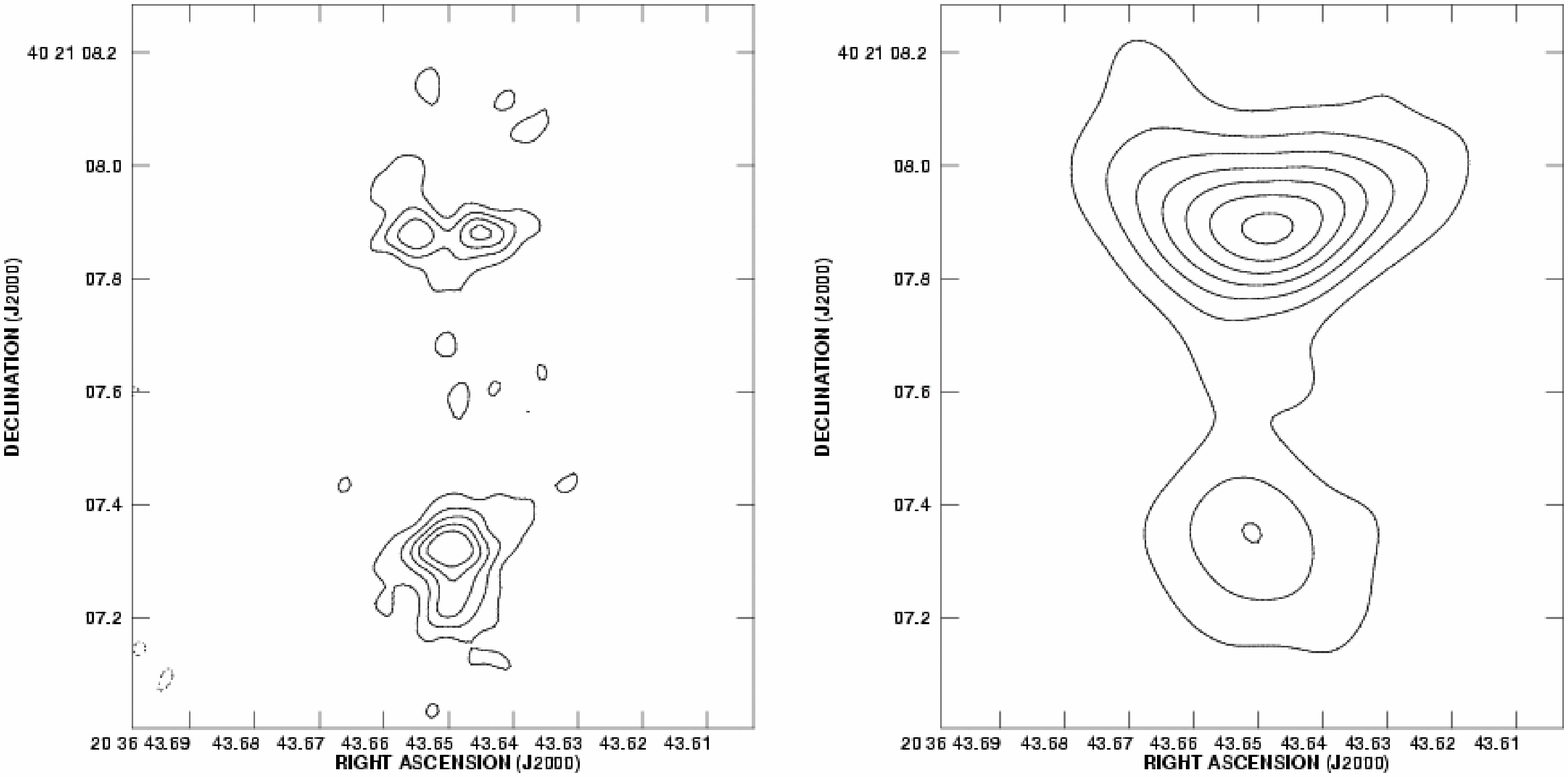}
\includegraphics[scale=0.55]{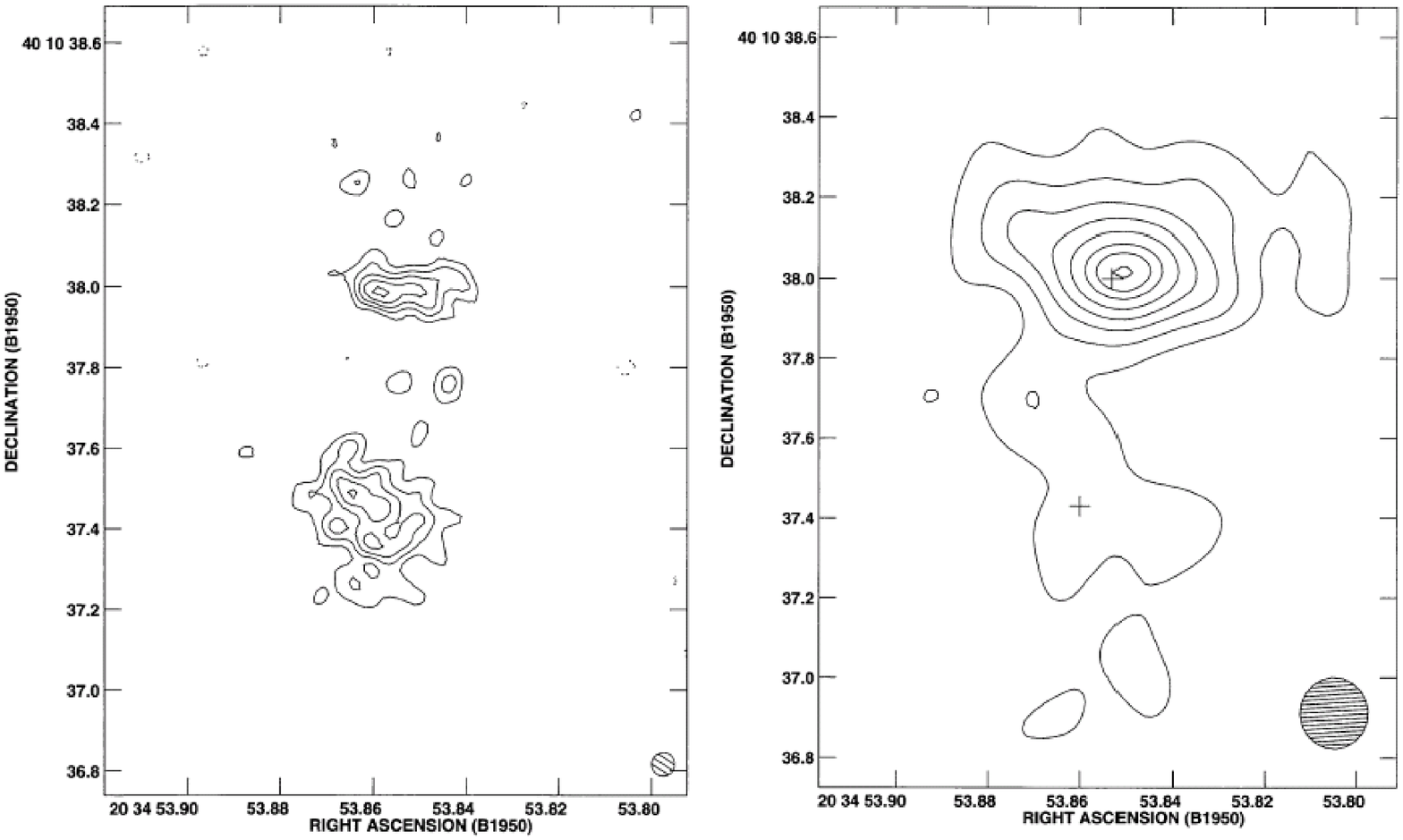}
\caption{\small Intensity distributions at 4.8 (left) and 1.6~GHz
(right) of our WR~147 model at $0^\circ $ inclination (top panels) and the
actual observations reproduced from \citet{Williams:1997} (bottom panels).  
The simulated MERLIN observations use the same u-v distribution and beam 
sizes as the actual observations in the bottom panels. The beams are circular, of diameter 57 and 175~mas 
at 4.8 and 1.6~GHz respectively. The contour levels in the simulated and actual
images are also the same. The similarity of the simulated and actual 
observations is striking.}
\label{fig:wr147_image}
\end{center}
\end{figure*}

To appreciate how the intensity distributions from the models
compare with the MERLIN observations shown in \citet{Williams:1997}, 
visibilities were generated from the model intensity distributions.
The resulting visibilities were
then imaged and deconvolved using the same procedure as in 
\citet{Williams:1997}, giving the ``simulated'' observations shown in the upper
panels of Fig.~\ref{fig:wr147_image}. The remarkable similarity between 
these images and the actual observations shown in the lower panels of
Fig.~\ref{fig:wr147_image} is striking. However, a critical eye reveals 
that the 5~GHz peak intensity of the collision region is a little lower 
than that of the WR stellar wind in the model, and the WR
star is a little too bright at 1.6~GHz. The WR star also appears to be
a little more compact than the observations show, and we
attribute these minor differences to our simple spherical, isothermal
model of the stellar winds. Note that the OB star is not visible in
the simulated observations. This is because the observations are
essentially the intensity distribution convolved
with the interferometer beam and the emission from the OB star is then
sufficiently dispersed that it is no longer visible.

Examination of the simulated images can help to
constrain the inclination angle. The synthetic 
observations shown in Fig.~\ref{fig:wr147_image} are
for an inclination of $0$\fdg\/ At an inclination angle of
$30^\circ$, the synchrotron emission is spread out over a much larger
area and has a much lower surface brightness than in the simulation
for $i = 0^\circ $, and in the actual observations. This fact, and the
spectral modeling, points to a fairly low value for the system inclination, 
between $0^\circ $ and $30^\circ$, though this is
contrary to the conclusions of \citet{Pittard:2002c} where larger
inclination angles were required to recover the extended X-ray
emission. This apparent contradiction may be resolved if the
stellar X-ray emission is extended on larger scales than previously
thought \citep[see][]{Skinner:2002}.

\section{Additional considerations and future perspectives}
\label{sec:other}
In this section we briefly review processes which, though important, 
have not been discussed elsewhere in this paper. We refer the interested
reader to \citet{Folini:2000} where some of these effects are discussed
in greater detail, including speculations on their mutual interaction. 
Theoretical considerations for clumped winds and 
the intriguing problem of dust formation are discussed in 
\citet{Folini:2000} and \citet{Walder:2002}.

\subsection{Radiation dynamics}
\label{sec:rad_dyn}
The influence of one star's radiation field on the dynamics of the
other star's wind has been quantified by \citet{Stevens:1994} and
\citeauthor{Gayley:1997} (1997, \citealp[see also][]{Pittard:1998}). 
In close binaries the winds may be inhibited in
their initial acceleration (since the {\em net} radiation flux is
reduced between the stars). In WR+O binaries, where the WCR occurs
closer to the O star than the WR star, the WR wind may be rapidly
braked if there is strong coupling with the O-star radiation
field. The opening angle of the WCR is a sensitive diagnostic of the
braking effect, although the geometry of the WCR is an ambiguous
constraint when the momentum ratio, $\eta$, of the winds is not known.
The reduction in pre-shock velocity and the change to the morphology
of the WCR have obvious consequences for the X-ray emission, but
should also impact the radio emission. 

\subsection{Orbit-induced curvature of the WCR}
\label{sec:large_scale}

Orbital motion will distort the structure of the WCR into a large-scale
spiral form, and is strikingly displayed in infrared observations of
WR~104 \citep{Tuthill:1999}. Theoretical models of the curvature range from
the analytical \citep{Canto:1999,Tuthill:2003}, to the numerical
\citep{Williams:1994}, to the hydrodynamical 
\citep{Pittard:1999,Walder:2003}.
The orbit-induced curvature of the WCR will affect the observed X-ray 
emission in close binaries \citep{Pittard:2002d} and the radio emission seen
from wider (especially eccentric) binaries. Theoretical simulations are
currently thin on the ground.

\subsection{Instabilities}
\label{sec:instab}
The WCR is subject to many different types of
instability. These include the classical Rayleigh-Taylor,
Kelvin-Helmholtz and Richtmyer-Meshkov instabilities which act on
the surface between the two winds, thin-shell instabilities which
act on isothermal shock-bounded slabs \citep{Vishniac:1994,Blondin:1996},
and the thermal instability from radiative cooling 
\citep[e.g.,][]{Strickland:1995}. Since these instabilities are not
isolated from one another, the resulting flow
may be very complex \citep[see, e.g.,][]{Stevens:1992}. While
the degree of instability in theoretical models may depend on assumptions made 
in the hydrodynamical code \citep[see][]{Myasnikov:1998b}, it seems
certain that the WCR will often be dynamically unstable. This may lead
to some X-ray \citep{Pittard:1997} and radio variability.

\subsection{Ionization equilibrium}
\label{sec:nei}
It is generally assumed that the shocked gas is in ionization
equilibrium, though in wide binaries there could be some departure from
this (see discussion in Henley et al. in these proceedings). The 
self-consistent inclusion of non-equilibrium ionization 
has yet to be incorporated into hydrodynamical models of CWBs. It will
be interesting to discover if future work in this direction is able to better
reproduce the X-ray line shifts seen in WR~140 where inconsistencies 
with current models exist.

\subsection{Ion-electron temperature}
\label{sec:ion_e_temp}
In wide binaries the shocks will be collisionless, and the immediate
post-shock electron and ion temperatures may differ. This possibility
has been investigated in a preliminary fashion by \citet{Zhekov:2000},
where the effect on the X-ray emission from the WCR in WR~140 was
examined. While the broad-band spectral fits to {\it ASCA} data 
indicated that a model with this effect was preferred, it is probably
too soon to draw strong conclusions from this work and 
further study is needed.

\subsection{Thermal conduction}
\label{sec:conduction}
There has been little work to date on thermal conduction in CWBs.
However, the high temperatures and the steep temperature gradients
in these systems indicate that thermal conduction may play an 
important role. Preliminary work has been carried out by 
\citet{Myasnikov:1998}, \citet{Motamen:1999}, and \citet{Zhekov:2003},
where it is shown that {\em efficient} thermal conduction 
increases the X-ray luminosity and softens its spectrum.
However, the presence of magnetic fields greatly reduces the thermal
conductivity in a direction perpendicular to the field lines, and this
has not yet been considered. This should be addressed in future
work since we know that magnetic fields are present in these systems 
due to the synchrotron emission which is observed.

\subsection{IC cooling of nonthermal particles}
\label{sec:iccool_nt}
While the simple models presented in \citet{Dougherty:2003} are able
to reproduce both the spectrum and the spatial distribution of radio
emission remarkably well, the neglect of IC cooling is a major
omission. To realistically model the observed radio emission from CWBs
the downstream evolution of the nonthermal electron energy spectrum
should be calculated.  This is being addressed in a follow-up paper
(Pittard et al., in preparation) where, in addition, the effect of 
Coulombic cooling
on the lower energy relativistic particles is also considered. 


\subsection{IC cooling of thermal particles}
\label{sec:iccool_thermal}
IC losses may be important not only for the nonthermal
shock-accelerated particles, but, at times, perhaps also for the hot
post-shock thermal gas \citep[see, e.g.,][]{White:1995b}.  More work
is needed in this area to determine when this might lead to a
significant change in the resulting thermal X-ray emission.

\section{Summary}
\label{sec:summary}
The wide range of parameter space covered by CWBs
means that they are useful probes of a wide range of physical
phenomena. X-ray observations in close binaries are a sensitive tool
to study radiative braking, while in wide binaries X-ray emission is a
useful diagnostic of the electron-ion temperature equilibration timescale.
The input of theoretical X-ray spectra into {\it XSPEC} via user generated
``table models'' has proved to be extremely useful in obtaining direct
constraints of the mass-loss rates of the binary components, and enjoys
several advantages over more traditional
techniques. Such work applied to $\eta$~Carinae may in future
determine if there is enhanced stellar mass-loss associated with the
periodic close approach of the stars.

On the other hand, radio observations can be used to explore the 
efficiency of diffuse shock acceleration at densities much higher than
those in other astronomical objects with high Mach number shocks, 
e.g., supernovae.
Preliminary work in this area indicates that it may be necessary to
consider the effect of particle acceleration on the shock
jump conditions. Another consequence of particle acceleration is that
subsequent IC cooling of the relativistic particles leads to
nonthermal emission at X-ray and $\gamma$-ray energies, e.g., 
\citet{Pollock:1987}, \citet{Benaglia:2003}. Such emission may already have
been detected at X-ray energies \citep{Rauw:2002}, and new and
forthcoming missions like {\it INTEGRAL} and {\it GLAST} may
observe this effect at $\gamma$-ray energies.

Future work should address these issues and those identified in
Sec.~\ref{sec:other}, and where possible radio and
X-ray data should be compared against theoretical predictions from a 
unified model. The review of (some) important processes and
mechanisms in Section~\ref{sec:other} demonstrates the richness of
physical effects occurring in colliding wind binaries which we
are striving to understand.

\section*{Acknowledgments}
We would like to thank the conference organizers for their invitation
and the excellent meeting which they put together. We have enjoyed
many stimulating and lively discussions with our colleagues over the years, 
and would like to thank them for collectively making our work
far more interesting.

\end{document}